\documentclass[aps,prl,twocolumn]{revtex4}
\usepackage{graphicx}\usepackage{color}
\begin{document}

\newcommand{\figureheight}{8.2 cm}
\newcommand{\putfig}[2]{\begin{figure}[h]
        \special{isoscale #1.bmp, \the\hsize \figureheight}
        \vspace{\figureheight}
        \caption{#2}
        \label{fig:#1}
        \end{figure}}

\newcommand{\eqn}[1]{(\ref{#1})}

\newcommand{\be}{\begin{equation}}
\newcommand{\ee}{\end{equation}}
\newcommand{\bea}{\begin{eqnarray}}
\newcommand{\eea}{\end{eqnarray}}
\newcommand{\bean}{\begin{eqnarray*}}
\newcommand{\eean}{\end{eqnarray*}}

\newcommand{\nn}{\nonumber}




\title{Influence of dimensionality on superconductivity in carbon nanotubes}
\author{{S. Bellucci}$^1$, M. Cini$^{1,2}$,  P. Onorato$^{1,3}$ and E. Perfetto$^{4}$ \\}
\address{
        $^1$INFN, Laboratori Nazionali di Frascati, P.O. Box 13, 00044 Frascati, Italy \\
        $^2$Dipartimento di Fisica,  Universit\`{a} di Roma Tor Vergata, Via della Ricerca Scientifica 1 00133,  Roma, Italy\\
$^3$Department of Physics "A. Volta", University of Pavia, Via Bassi 6, I-27100 Pavia, Italy\\
$^4$Consorzio Nazionale Interuniversitario per Le Scienze Fisiche
della Materia, Universit\`{a} di Roma Tor Vergata, Via della
Ricerca Scientifica 1 00133,  Roma, Italy}
\date{\today}

\begin{abstract}
We investigate the electronic instabilities in carbon nanotubes
(CNs), looking for the break-down of the one dimensional Luttinger
liquid regime due to the strong screening of the long-range part of
the Coulomb repulsion. We show that such a breakdown is realized
both in ultra-small single wall CNs and multi wall CNs, while  a
purely electronic mechanism could  explain  the superconductivity
(SC) observed
 recently in   ultra-small  (diameter $ \sim
0.4 nm$) single wall CNs ($T_c\sim 15\; ^{o}K$)  and  entirely
end-bonded multi-walled ones ($T_c\sim 12 \;^{o}K$). We show that
both the doping and the screening of long-range part of the
electron-electron  repulsion, needed to allow the SC phase, are
related to the intrinsically 3D nature of the environment where the
CNs operate.

\end{abstract}

\maketitle

{\it Introduction --} The recent progresses in nanotechnology
allowed for a detailed study of the transport properties of 1D
electron systems.
The discovery of carbon nanotubes (CNs) in 1991\cite{1}, as a
by-product of carbon fullerene production, opened a new field of
research in mesoscopic physics\cite{cbmac} expecially because of
their potential application to nanoelectronic devices. It is known
that two types of CNs, i.e. single wall carbon nanotubes (SWNTs)
and multi wall carbon nanotubes (MWNTs) (see Fig.1), exist and are
reported to display different electronic properties depending on
their diameter and on the helicity of the carbon rings around the
tubule\cite{saito}. Because of their sizes CNs usually  behave as
ideal one dimensional (1D) electronic systems and there have been
many experiments showing the existence of superconducting (SC)
correlations in these devices at low temperatures.
\begin{figure}
\includegraphics*[width=1.0\linewidth]{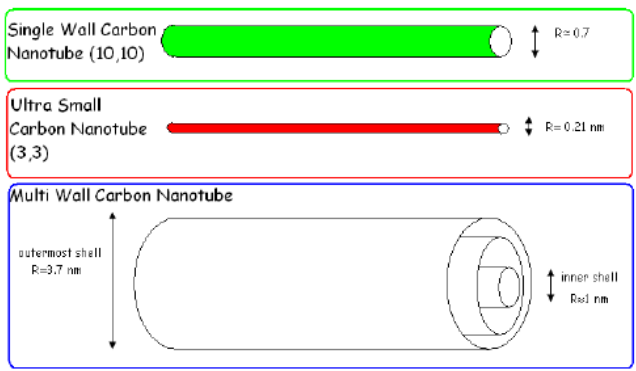}
 \caption{{A SWNT is a  rolled up sheet of graphite just nanometres in
diameter and length up to some microns, while a MWNT is made by
several concentrically arranged graphene sheets with a radius of
some nanometers.}}
\end{figure}

The SC behaviour in low dimensional systems is a quite interesting
question since 40 years ago
 Mermin and Wagner\cite{MW} proved a famous theorem stating that it is impossible
for abrupt phase transitions with long-range order to occur in 1-
or 2-D systems at finite temperature. Thus CNs are among the best
candidates for investigating the possibility of (quasi)1D
superconductivity. In general CNs do not show superconducting
properties but some recent experiments found that ultra-small
single wall carbon nanotubes\cite{[11]}  and entirely end-bonded
multi-walled ones\cite{tk} can superconduct. Clear evidence of
superconductivity was  also found in CNs suspended between
superconducting contacts, showing the so-called proximity
effect\cite{7,8} while genuine superconducting transitions below
$1^oK$ have been observed in thick ropes of nanotubes suspended
between normal and highly transparent electrodes\cite{10b}.

Here we mainly discuss the phenomenology concerning the
experiments of ref.\cite{[11]} and ref.\cite{tk}. In the first
 experiment\cite{[11]}, ultra-small-diameter single wall
nanotubes (USCN) have been produced  inside the channels of a
zeolite matrix (with inner diameter of  $\sim 0.73\; nm$). The
nanotube diameter $d = 4.2\pm0.2\AA$  is closer to the value
calculated for a $(3,3)$ CN geometry, although the presence of (5,0)
nanotubes cannot be discarded\cite{[14]}. These CNs have many
unusual properties, such as superconductivity, leading to a
transition temperature $T_c \approx 15^o K$\cite{[11]}, much higher
than that observed in bundles of larger diameter tubes \cite{ropes}.

In a recent letter\cite{tk} it was reported that there is a
superconducting phase competing with the Luttinger liquid (LL)
phase and even overcome it in entirely end-bonded MWNTs  with a
transition temperature $T_c$ as high as $12^o K$. The lengths of
the MWNTs were $L\sim 0.6 \mu m$ while the high-resolution
cross-sectional TEM images showed
 a MWNT with an outer diameter of $2R_{o}=7.4 nm$, and  inner diameter
of $2R_{i}< 2 nm$ ($R< 1nm$). There was  also found  that the
emergence of this superconductivity is highly sensitive to the
junction structures of the Au electrode/MWNTs.  $T_c$ depends on the
numbers of electrically activated
 shells; to enhance superconductivity the
Au electrodes  must be in contact  with the tips of all the shells;
in contrast,  the  conventional "bulk junction" contacts   touch
only the outermost shell of a tube.
  Below $T_c$ the LL states are suppressed and an SC behaviour
 can appear while for $T > T_c$  a conventional LL behaviour was observed.

In this paper we investigate theoretically this phenomenology by
focusing on the central role which the screening of the long-range
part of the Coulomb repulsion plays. This screening has to be
generally related to the environment (intra- and inter-shell
screening in the MWNTs or screening by the zeolite matrix for the
USCNs) and has an intrinsic three dimensional nature. Thus our aim
is to show that the interplay between the 1D typical character of
the CNs and the 3D nature of the environment allows the SC phase.

In order to pursue our aim  we discuss  the possibility that a
superconducting behaviour can arise in these CNs
 by a purely electronic mechanism, i.e. neglecting the contribution
of phonons. We do that  by a comparison between two different
approaches, one  based on the Luttinger model the other one, which
emphasizes the role of the lattice and short range interaction,
developed starting from the Hubbard Hamiltonian.

 \

{\it Transport in 1D electron systems --} Electronic correlations
have been predicted to dominate the characteristic features in quasi
1D interacting electron systems, leading to the breakdown of the
conventional Fermi liquid picture. In fact Landau quasiparticles are
unstable in 1D and the low-energy excitations take the form of
plasmons (collective electron-hole pair modes). Thus the 1D
character of the system leads to a strong correlation among
electrons, inducing of the so-called Luttinger liquid
(LL)\cite{lutt}.  The LL state has two main features:

i) the power-law dependence of physical quantities, such as the
density of states (DOS), as a function of energy or temperature;

ii) the spin-charge separation: an additional electron in the LL
decays into decoupled spin and charge wave packets, with different
velocities for charge and spin.

Characteristic experimental signatures support the assumed LL
behaviour of CNs\cite{ll,ll2,17}, where the temperature dependence
of the resistance above a crossover temperature $T_c$ was
measured\cite{Fischer}. In fact the power-law dependence of physical
observables follows from the behavior of the DOS as a function of
the energy. For example, the tunneling conductance $G$  in  a small
bias experiment\cite{kf} follows a power law,
 \bea
 G=dI/dV\propto
T^{\alpha } \eea  for $eV_b\ll k_BT$, where $V_b$ is the bias
voltage, $T$ is the temperature and $k_B$ is Boltzmann's constant.
The critical exponent $\alpha$ assumes different values   for an
electrode-bulk junction ($\alpha_{bulk}$) and for an electrode-end
junction ($\alpha_{end}$), as  reported for MWNTs in
Ref.\onlinecite{1b}.

\

The theoretical analysis start from the model of the CN where  the
electrons have linear dispersion relation around each of the two
Fermi points  at $(\pm K_F,0)$ ($K_F=4\pi/3a$, and $a=2.46${\AA}
is the lattice constant). These branches are highly linear with
Fermi velocity $v_F\approx 8\times 10^5$ m/s. The linear
dispersion relation holds for energy scales $E < D$, with the
bandwidth cutoff scale $D\approx \hbar v_F/R$ for tube radius $R$.

\

For what concerns the interaction we distinguish some processes
associated with the Fermi points $\pm K_F$, (a) the {\it forward
scattering} ($g_2$ with small transferred momentum $q \sim
q_c=2\pi/L$  which can be assumed as the natural infrared cut-off,
depending on the longitudinal length $L$ of the CN) (b) the {\it
backscattering} ($g_1$ with large transferred momentum i.e. $q\sim 2
K_F$), (c) an additional {\it Umklapp} process which is relevant at
half-filling and  in our case we neglect, since the sample is
assumed to be doped (d) an additional {\it forward scattering} term
($f$) which
 measures the difference between intra- and inter-sublattice
interactions; this term is  due to the hard core of the Coulomb
interaction.i.e. it follows from the unscreened short range
component of the interaction. $f$ corresponds to\cite{lutt}
$\delta V_p= U_{++}-U{+-}$, where $U_{p,p'}$ is the interaction
between electrons belonging to different sublattices ($p,p'$), and
it is strongly suppressed at a distance much larger than $\ell\sim
0.3 nm$\cite{noijpcm}. In the same way, the only non-vanishing
contribution to $g_1$ comes from $|x-x'|\leq a$, because of
rapidly oscillating contributions\cite{lutt}. Thus we can assume
$g_2$ as the only relevant long range component of the
interaction. Moreover notice that the CN's dimensions play a
central role in determining the strength of the different terms of
the interaction. The radius $R$ and the length $L$ yield two
natural cutoffs, $\approx \frac{2\pi}{R}$ and $q_c$ while we can
classify the physical quantity according their dependence on the
radius. Thus we have a long range  coupling, $g_2= \hat{V}_0(q_c)$
weakly dependent on the radius, and two strong dependent
interaction couplings, $g_1 = \hat{V}_0(2 K_F)$ and $f$ which
scale as $1/R$[\onlinecite{lutt}].

 After introducing the values of the couplings the effective field theory can be  solved in
practically exact way  as was done in  ref.\cite{lutt} and the
bulk critical exponent  has the form\cite{noi,npb,prb}
\begin{equation}\label{al1}
  \alpha_{bulk}=\frac{1}{4} \left(g+\frac{1}{g}-2 \right),
\end{equation}
where $g$ depends just on the Forward scattering as \bea\label{K}
 \sqrt{1 + \frac{g_2}{ (2
\pi  {v}_F)}} =\frac{1}{g}. \eea It follows that  usually the short
range terms of the interaction can be neglected. However it was
predicted\cite{lutt}  that the effects of $g_1$ and $f$ can be
dominant  at  low temperatures ($T$ below the crossover
temperatures,  $g_1\rightarrow k T_b=  D e^{-\frac{2\pi v_F}{g_1}} $
and $f\rightarrow k T_f=  D e^{-\frac{2\pi v_F}{f}}\lesssim k T_b$ )
where the Luttinger liquid breaks down and a (quasi-) long-range
order phase appears. For long-ranged interactions (which is the case
of CNs in typical conditions), we have $T_f \sim T_b$, while for
short-ranged interactions it results $T_f < T_b$. In the latter case
a superconducting instability is predicted at $T \sim T_f$ if the
Luttinger liquid parameter $g$ is larger than $1/2$.

\

{\it Breakdown of the Luttinger Liquid --} The above discussion
shows that a pure electronic mechanism which gives
superconductivity according the LL theory needs  a screening of
the forward scattering, $g_2$  (long range effect $g>0.5$), an
increase of the backward scattering, $g_1$ (short range effect
$T_b$) and aid from the $f$ scattering (high value of the
corresponding temperature, $T_f$).

In typical isolated  CNs of $R\gtrsim 1 nm$ no SC behaviour was
observed. This is in line with the above discussion, since  $g_1$
and $f$  are small compared to $g_2$ and  the estimated transition
temperature
  $T_f \sim T_b \sim 1mK$,
is very low indeed. In this paper  we argue about the drastic
effects that one can achieve  1) by the interaction with other
tubes and/or a matrix   2) by  an ultra-small CN radius 3) by
doping. The interactions with the surroundings can provide an
effective screening of the long range component of the interaction
and strongly enhance $T_c$ . This ia an  interplay between   3D
effects and the quasi 1D behaviour that can allow  a SC
transition; further 3D effects will be pointed out  below for the
end bonded MWNTs. The ultra-small radius contributes to enhance
the short-range interaction. Although at first sight a stronger
$g_{1}$ might seem to make pairing more efficient, in fact the
converse is found to be true both in the LL approach and in the
Hubbard model (see below). The effects of doping will also be
discussed below.

In fact in the experimental samples of Ref.\onlinecite{[11]} the
CNs are arranged in large arrays with triangular geometry (with
intertube distance $d \approx 1$ nm), behaving as a genuine 3D
system concerning the screening properties. The presence of many
CNs inside the zeolite matrix provides
 a large reduction of $g_{2}$ (by a
factor $\approx 10^{-2}$),  while the short range components have
to remain almost unchanged. This allows for the occurrence of a
sizable superconducting instability within the Luttinger liquid
approach.

\begin{figure}
\includegraphics*[width=1.0\linewidth]{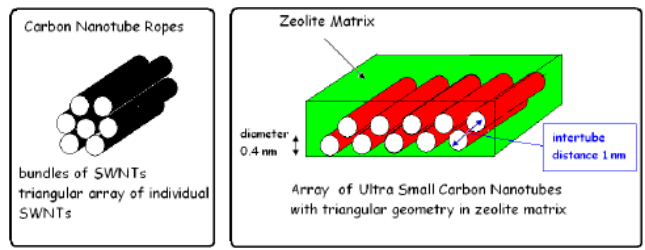}
 \caption{{As already pointed out in Refs.\onlinecite{gonzperf} and
\onlinecite{gonzperf2}, the intra-tube Coulomb repulsion at small
momentum transfer (i.e. in the forward scattering channel) is
efficiently screened by the presence of electronic currents in
neighboring nanotubes.}}
\end{figure}

\

The SC transition in end-bonded MWNTs can be also  explained as an
effect of the interplay between a 3D and a 1D behaviour. In fact
the power-law of the conductance observed for $T>12^oK$ in the
MWNTs of ref.\cite{tk} is consistent with the Luttinger liquid
character of the normal state. Therefore the observed sharp
breakdown of the power-law at $12^oK$ is an indication that an
approach based on the superconducting
 instability of the Luttinger liquid is well posed.

 The experimental  results are
consistent with the idea that only the outermost shell became
electrically active in the Au-bulk junction (see Fig.3). In this
case the conducting channel is not efficiently screened and
retains a strong 1D character. The entire end-bonding of MWNTs
made all the shells electrically active, while only some of the
shells were electrically active in the partial Au-end junctions.
Therefore, when the electrical contacts touch the top of all the
shells (as in the case of the entirely end-bonded MWNTs), the
activation of the internal shells gives a large dielectric effect,
due to intra- and inter-shell screening, and at the same time it
provides an incipient 3D character, which is crucial for
establishing the superconducting coherence\cite{perfgonz}.

Thus we hypothesized  that all contacted shells can transport the
normal current as resistors in parallel connection. In the normal
(Luttinger) state, a current  of electrons flows in each shell
while  the conductance $G$ is mainly given by the outermost
shells, because the outermost shells have the smallest resistance.
When the temperature decreases below  $T_c$ the  superconductivity
can be  favored in the inner shells  of the MWNT because the short
range interactions are enhanced when
 the radius decreases. The long range term of the interaction, $g_2$, in the end-bonded MWNTs is screened by the
electronic currents located in the sorrounding  shells. Notice
that also in this case the screening of $g_2$ is essentially due
to the 3D nature of the MWNTs.  An analogous discussion could be
extended to the ropes of CNs analyzed  in  ref.\cite{10b}.
\begin{figure}
\includegraphics*[width=1.0\linewidth]{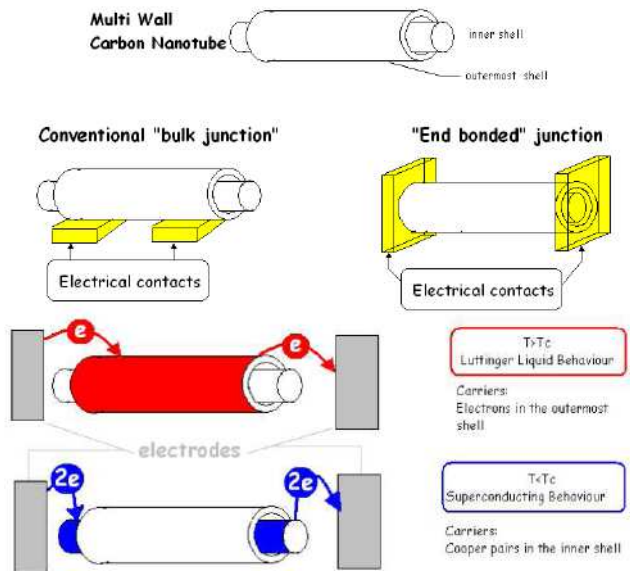}
 \caption{
{Electrical contacts in the systems of ref.\cite{tk}   made of
$Au$ are bonded to the tubes so they touch the top of all the
shells while  conventional  "bulk junction" contacts, in contrast,
touch only the outermost shell of a tube and along its length. In
usual conditions, transport measurements carried out in MWNTs
reflect the electronic properties of the outer shell, which the
electrodes are attached to. On the other hand, in entirely
end-bonded samples the inner shells are electrically active, with
relevant consequences. In particular the innermost one is able to
support the transport of Cooper pairs below a temperature
consistent with the measured one. }}
\end{figure}

\

{\it Beyond the Luttinger Model --} The above discussion  provides
a consistent explanation of the presence of a phase quite
different from the normal (LL) one. We could also predict the
crossover temperatures as reported above by using the results of
ref.\cite{lutt}. In this way we obtain values of $T_c$ that are
compatible with the experimental findings but are quite
inaccurate. In order to improve our predictions we develop a
different approach based on the Hubbard model which emphasizes the
role of the lattice and short range interaction.

A similar analysis was developed in ref.\onlinecite{krot} where the
superconductivity in CNs  was investigated with the renormalization
group technique. By introducing the parameter calculated for the CNs
analyzed here in the model of ref.\cite{krot} we find that a SC
phase is supported  just in the presence of a slight doping.
Unfortunately  also in this approach the  estimate of $T_c$ is quite
inaccurate. A Hubbard like approach which gives a prediction about
the critical temperature was proposed in ref.\cite{noijpcm}. It is a
pure electronic mechanism which leads to superconducting pairing
starting from the Hubbard model on the wrapped honeycomb lattice
away from half filling\cite{psc}. In this theory the ultra-small
radius CN are favored for two reasons: first, the Hubbard on-site
repulsion $U$ is larger for smaller radius, and the pairing energy
$\Delta$ in thereby enhanced, as we already recalled; second,  at
fixed $U,$ one finds that $\Delta$ increases with decreasing radius.
By using the BCS formula $\Delta = 1.76 kT_c$ for the mean field
transition temperature   for the Ultra Small CNs of ref.\cite{[11]}
one estimates $T_c\approx 7 \div 70 ^oK $ which is compatible with
the measured one. We observe that the lower bound $T_c \sim 7K$
takes into account that $\Delta$ may vary of about one order of
magnitude away from optimal doping. Although the theory of
ref.\cite{noijpcm} has not yet been extended to the MWNT geometry,
we may expect that the inter-shell hopping should slightly enhance
$\Delta$ compared to the single-shell case. Thus, we can roughly
estimate the crossover temperature for the MWNTs of ref.\cite{tk}.
At optimal doping we obtain $ T_c\approx 4 \div 40 ^oK$ This value
is slightly lower than the one of USNTs, in qualitative agreement
with the experimental findings. Also in this case the lower bound
$T_c \sim 4^oK$ is understood in terms of possible deviation from
optimal doping \cite{noiprb07}.

\

{\it Discussion --}Here we want to discuss the intrinsic 3D nature
of the model and the most relevant effects that we have to take in
account.

The Hubbard model keeps only the on-site repulsion $U$  and  can
be safely used under the condition that the long range component
of the e-e interaction is well screened. Such screening  is due to
the genuine 3D nature of the systems under study.


The presence of doping deserves a specific discussion and it has
to be related to some kind of external effect.

MWNTs use to be significantly doped, what leads to the presence of
a large number of subbands at the Fermi level \cite{[10]}. The
contribution of a large number of modes at low energies has then
an appreciable impact in the enhancement of observables like the
DOS while the activation of several channels is also responsible
of the $g_2$ screening. This topic was investigated in
refs.\cite{npb,prb} where  the effect of doping in the suppression
of tunneling observed in MWNTs was studied. There was shown how
the doping is related to disappearance of the typical 1D behaviour
also by modifying the effective dimensionality of the system. The
doping induced crossover from a 1D to a 3D behaviour is analogous
to the transition from the LL to SC phase.

 For what concerns the doping in the SWNTs of ref\cite{[11]} the
 presence of the 3D environment, surrounding matrix or nearest
 CNs, can be assumed as the main cause of the doping.
  Despite the differences, the effects of the doping (screening, dimensional
 crossover) are quite similar to the ones discussed in the case of
 MWNT.

 Thus we can conclude that the doping not only  plays a central role in the SC instability in
 the Hubbard model, but contributes to
 the breakdown of the strictly 1D behaviour by adding several 1D
 conducting channels at the Fermi level and supporting the
 screening of the long range interaction.

S.B. and P.O. acknowledge partial financial support from the grant
2006 PRIN "Sistemi Quantistici Macroscopici-Aspetti Fondamentali ed
Applicazioni di strutture Josephson Non Convenzionali''.




\bibliographystyle{prsty} 

\bibliography{}

\end{document}